\documentclass[iop]{emulateapj}

\usepackage{natbib}
\usepackage{enumerate}
\usepackage{graphicx}
\usepackage{xfrac}
\usepackage{hyperref}

\usepackage{color}

\begin{document}
\title{Variable H$^{13}$CO$^+$ Emission in the IM Lup Disk: X-ray Driven Time-Dependent Chemistry?}
\shorttitle{Variable Ion Chemistry in a Young Circumstellar Disk}
    \shortauthors{Cleeves et al.}

 \author{L. Ilsedore Cleeves\altaffilmark{1,2}, Edwin A. Bergin\altaffilmark{3}, Karin I. {\"O}berg\altaffilmark{1},  Sean M. Andrews\altaffilmark{1}, David J. Wilner\altaffilmark{1}, Ryan A. Loomis\altaffilmark{1}}

\altaffiltext{1}{Harvard-Smithsonian Center for Astrophysics, 60 Garden Street, Cambridge, MA 02138}
\altaffiltext{2}{Hubble Fellow} 
\altaffiltext{3}{Department of Astronomy, University of Michigan, 1085 S. University Ave., Ann Arbor, MI 48109}

\begin{abstract}
We report the first detection of a substantial brightening event in an isotopologue of a key molecular ion, HCO$^+$, within a protoplanetary disk of a T Tauri star. The H$^{13}$CO$^+$ $J=3-2$ rotational transition was observed three times toward IM Lup between July 2014 and May 2015 with the Atacama Large Millimeter Array. The first two observations show similar spectrally integrated line and continuum fluxes, while the third observation shows a doubling in the disk integrated $J=3-2$ line flux compared to the continuum, which does not change between the three epochs. We explore models of an X-ray active star irradiating the disk via stellar flares, and find that the optically thin H$^{13}$CO$^+$ emission variation can potentially be  explained via X-ray driven chemistry temporarily enhancing the HCO$^+$ abundance in the upper layers of the disk atmosphere during large or prolonged flaring events. If the HCO$^+$ enhancement is indeed caused by a X-ray flare, future observations should be able to spatially resolve these events and potentially enable us to watch the chemical aftermath of the high-energy stellar radiation propagating across the face of protoplanetary disks, providing a new pathway to explore ionization physics and chemistry, including electron density, in disks.
\end{abstract}


\section{Introduction}\label{sec:intro}
The gas-phase abundances of key molecular ions, including HCO$^+$, N$_2$H$^+$, DCO$^+$, and H$_2$D$^+$, are sensitive to high energy (keV -- GeV) processes in the dense interstellar medium \citep[e.g.,][]{caselli1998,williams1998,caselli2002,dalgarno2006}. In protoplanetary disks, abundances of such molecular ions, especially HCO$^+$ \citep{cleeves2014par}, are strongly influenced by X-rays from their host star \citep{glassgold2012}.

The X-ray emission from young stars is thought to originate from a combination of magnetic/coronal activity and/or accretion with plasma temperatures between a few MK to many tens of MK \citep[e.g.,][]{feigelson1999,kastner2002,pe2005,preibisch2005}. Typical source luminosities are substantial, ranging between $L_{\rm X}\sim10^{29}-10^{31}$ erg s$^{-1}$, and are often time variable \citep{feigelson1999}. Variability can occur both on long ($>>$ years)  and short (hours -- days) time scales. Longer term, gradual variability is attributed to the global evolution of the stellar magnetic field topology and/or disk accretion properties \citep[e.g.,][]{feigelson1999,pe2005,stelzer2015}. Shorter term variability is associated with magnetic reconnection events in the corona, similar to solar X-ray flares but with greater magnitude and frequency \citep[e.g.,][]{feigelson2007}.  Such events can result in luminosity changes up to two orders of magnitude within a matter of hours.

Given the sensitivity of molecular ions to energetic processes, it has been an open question whether the bulk chemistry of the disk responds to stellar X-ray variability \citep{ilgner2006,cleeves2015tw}. In this Letter, we present new multi-epoch {\em Swift} X-ray observations and ALMA observations of an optically thin isotopologue of HCO$^+$ toward the young solar-mass star, IM Lup. The ALMA observations reveal the first evidence of significant short-term chemical variability in the bulk molecular disk in an X-ray sensitive molecule. To explore one possible explanation for this newly discovered variability, we carry out simple models of time evolving X-ray irradiation of circumstellar disk material to specify the kinds of energetic events that may be responsible for the observed chemical behavior.

\section{Observations}\label{sec:obs}
\subsection{ALMA Observations}\label{sec:almatest}

\begin{deluxetable*}{ccccc}[hb!]
\tablecolumns{5}
\tablewidth{0pt}
\tablecaption{H$^{13}$CO$^{+}$ $J=3-2$ Observations \label{tab:obs}}
\tabletypesize{\footnotesize}
\tablehead{{Date} & Min Baseline & Max Baseline & Disk-Integrated &  Disk-Integrated \\
                       & &        &  H$^{13}$CO$^+$ $J=3-2$ & 1.15 mm Continuum}
\startdata
07/17/2014 & 20.1 m&650.3 m  & $0.466\pm0.032$ Jy km s$^{-1}$ & $0.286\pm0.003$ Jy \\
01/29/2015 & 15.1 m&348.5  m  & $0.590\pm 0.020$ Jy km s$^{-1}$& $0.308 \pm 0.001$ Jy \\
05/13/2015 &21.4 m&558.2 m & $1.254 \pm 0.019$ Jy km s$^{-1}$&  $0.307 \pm 0.001$ Jy
\enddata
\end{deluxetable*}

Observations of IM Lup in H$^{13}$CO$^+$ $J=3-2$ were carried out with ALMA as part of two separate programs, project codes ADS/JAO.ALMA\#2013.1.00226 (PI: {\"O}berg) and ADS/JAO.ALMA\#2013.00694 (PI: Cleeves). The 2013.1.00226 data were taken on 17 July 2014 with 32 antennas; the analysis was previously reported by \citet{oberg2015im}, and will not be repeated here. The 2013.00694 observations were taken on 29 January 2015 (40 antennas) and 13 May 2015 (36 antennas). The time spent on source was 21, 17, and 34 minutes for the 17 July 2014, 29 January 2015, and 13 May 2015 observations, respectively. The latter two observations were processed by the ALMA/NAASC staff; the quasar J1517-2422 was used to calibrate the bandpass, J1610-3958 was used to calibrate the phases, and Titan set the amplitude scale. The 2013.1.00226 data were self calibrated as described in \citet{oberg2015im} and the 2013.00694 data were both self-calibrated with three rounds of phase calibration on the disk continuum emission. The observing parameters are presented in Table~\ref{tab:obs}. 

\begin{figure*} 
\begin{centering}
\includegraphics[width=1.0\textwidth]{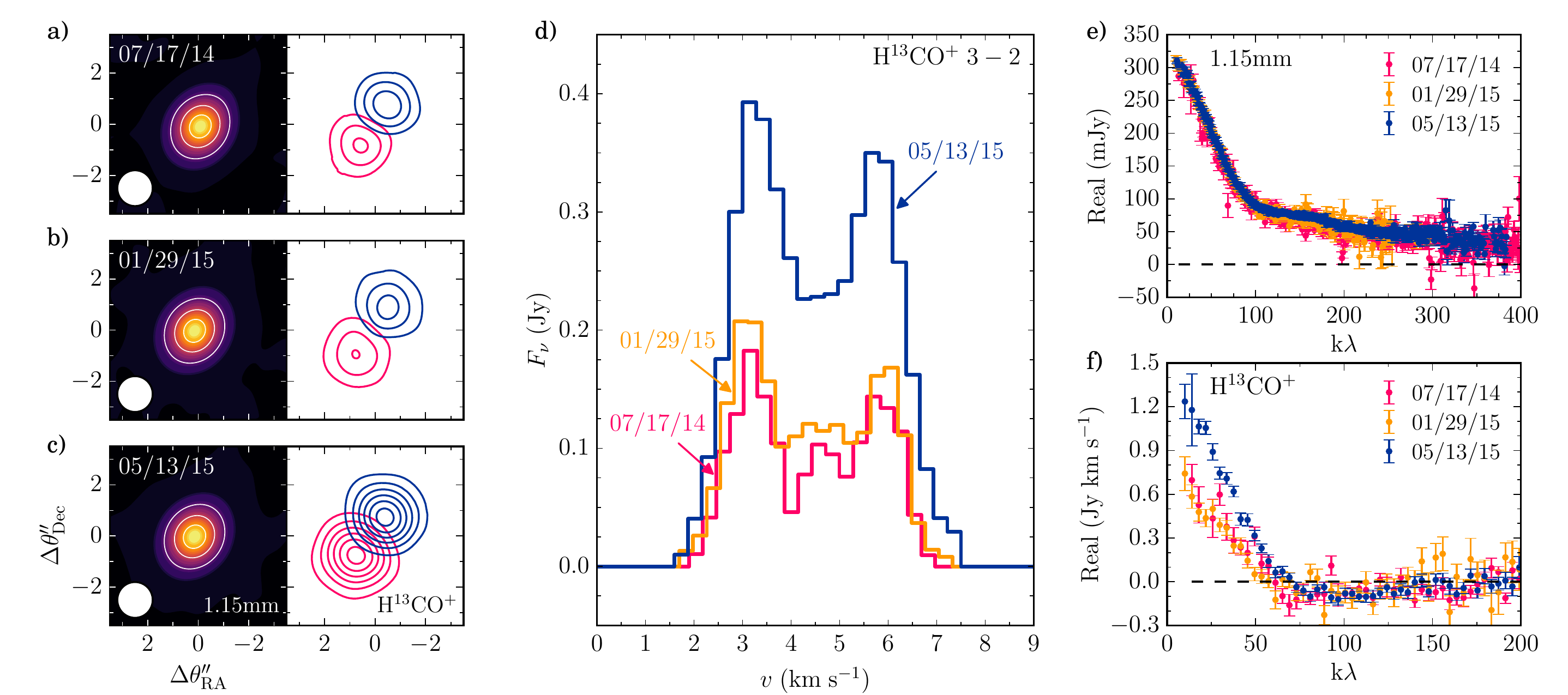}
\caption{a-c) 1.15 mm continuum (left) and velocity integrated line intensity of H$^{13}$CO$^+$ $J=3-2$ (right) in the image plane for the three dated observations. Line and continuum images are restored to the same $1\farcs4$ beam. The continuum contours (overlaid in white) are 50, 100, and 150 mJy. The line contours are shown for the red ($4.9-7.1$ km s$^{-1}$) and blue ($1.8-4.0$ km s$^{-1}$) shifted emission. Line contours start at 40 mJy per beam km s$^{-1}$ and increase by 40 mJy per beam km s$^{-1}$ intervals. d) Disk-integrated, continuum-subtracted spectra of H$^{13}$CO$^+$ $J=3-2$. e-f) Real visibilities for the 1.15mm continuum (top) and line integrated over the central 16 channels, $2.4-6.6$ km s$^{-1}$ (bottom). \label{fig:alma}}
\end{centering}
\end{figure*}

As the H$^{13}$CO$^+$ $J=3-2$ observations were taken at different times from different programs, the UV coverage and spectral configuration have slight differences. We channel-averaged the 17 July 2014 observations (61.035 kHz / 0.07 km s$^{-1}$) to match the spectral resolution of the latter two datasets, 244.141 kHz or 0.28 km s$^{-1}$ with the CASA task \texttt{cvel}. 
To correct for the beam differences, we imaged the data with \texttt{clean} using a UV taper and natural weighting to produce a circular $1\farcs4$ beam.

Disk-integrated spectra were extracted from the image cube with the same mask used in \texttt{clean}, which traces the gas motion. The RMS uncertainty on the integrated flux was estimated with the same mask applied to line-free frequencies. The disk-integrated continuum was calculated in the same spectral window containing H$^{13}$CO$^+$ $J=3-2$ within an elliptical annulus. The RMS uncertainty on the continuum was estimated with the same annulus off source. The line and continuum fluxes and RMS uncertainties are reported in Table~\ref{tab:obs}. 

The integrated line and continuum fluxes are consistent between the 17 July 2014 and 29 January 2015 observations.
The third observation, 13 May 2015, displays a jump in integrated line flux for H$^{13}$CO$^+$ $J=3-2$ while the continuum remains constant compared to earlier observations. Figure~\ref{fig:alma} presents images, spatially integrated spectra, and spectrally integrated line and continuum visibilities. The change in H$^{13}$CO$^+$ $J=3-2$ is visible not only in the data products but also directly in the visibilities before and after continuum subtraction. 

To ensure the variability is robust, we ran a number of checks on the data, examining the individual scans and the calibrators' phase and amplitude. Flux calibration should also not be an issue as we are comparing the line-to-continuum ratio. 
We also investigated the effects of different UV coverages. From the data, we created a ``model'' H$^{13}$CO$^+$ $J=3-2$ cube and a continuum image that is smoothed and clipped of noise. We sampled the model with the same spatial frequencies of the ALMA observations using \texttt{vis\_sample}\footnote{\url{https://github.com/AstroChem/vis_sample}.}. This test allows us to estimate the effect of spatial filtering for a model of known flux. The 17 July 2014 data are the most sparsely sampled at short baselines, retrieving a $\sim10\%$ lower line and continuum flux compared to the 29 January 2015 and 13 May 2015 configurations, which retrieve essentially identical fluxes to within 0.1\%. Given the difference in short spacings, we repeat the test with only UV distances of $\ge35$k$\lambda$, i.e., where all observations are well-sampled. The 17 July 2014 observations still return lower fluxes but at a reduced level, 6\% and 9\% fainter in the line and continuum, respectively.

This systematic effect introduced by the UV coverage is larger than the RMS uncertainty (Table~\ref{tab:obs}), and so we have chosen to present the line-to-continuum ratios in two ways. First, we compare the ratio measured for all three dates including both RMS and systematic (UV sampling) uncertainties for spatial frequencies $>35$k$\lambda$. Second, we compare just the 29 January 2015 and 13 May 2015 observations with similar UV coverage, whose uncertainty is RMS noise dominated. 

\subsection{Swift XRT and UVOT Observations}\label{sec:swift}
A series of {\it Swift} \citep{gehrels2004} X-ray and FUV observations of IM Lup were obtained through University of Michigan's {\it Swift} allocation and augmented by Target of Opportunity observations. Monitoring was carried out during 27 July 2014 - 28 August 2013, 14 January 2015 - 4 February 2015, and 2 August 2015 - 8 August 2016. The observations accumulated a total of 32.9 ks of XRT data in Photon Counting mode and 11.4 ks of UVOT in the UVM2 filter. The X-ray spectra were extracted using the online UK Swift Science Data Centre pipeline\footnote{\url{http://www.swift.ac.uk/user_objects/}} \citep{evans2009}. The photometric UVOT measurements were extracted using HEASoft with the \texttt{uvotsource} routine to compute the flux at 2246~\AA\ (the UVM2 filter). 

\begin{figure}[ht!]
\begin{centering}
\includegraphics[width=0.46\textwidth]{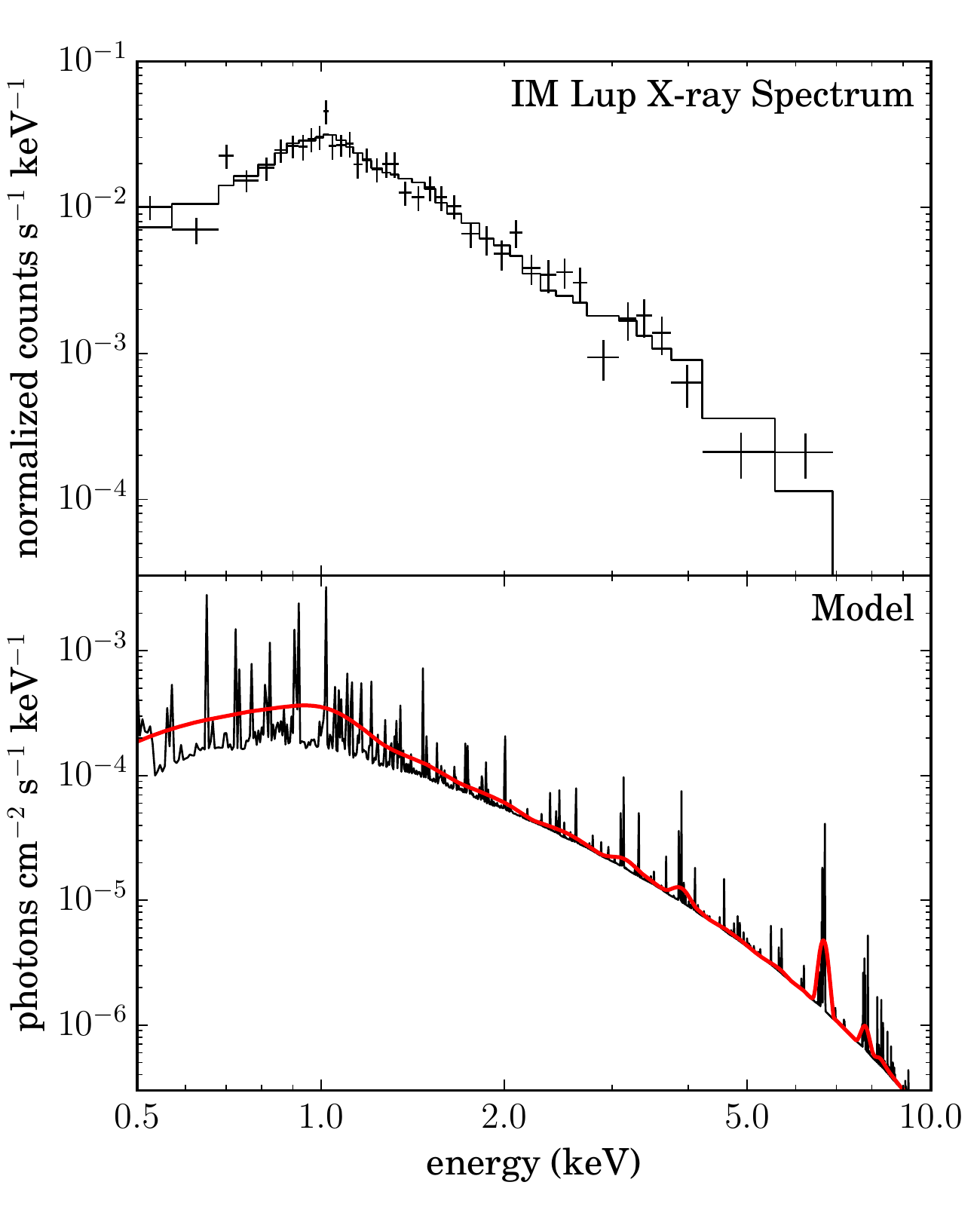}
\caption{Top: {\em Swift} spectrum of IM Lup including all epochs with the best fit folded model overlaid in solid black. Bottom: model spectrum at full resolution (black) and smoothed (red), where the latter is used as the input source spectrum pre-propagation through the disk. \label{fig:xspectrum}}
\end{centering}
\end{figure}

The X-ray spectra were modeled with HEASoft assuming \citet{grsa} abundances and three plasma components identified in \citet{gunther2010} at 0.3, 0.6 and 2.1 keV. In fitting the {\em Swift} data, we fixed the temperatures of these three components and allow their normalizations and absorption column densities to vary. We fit both the individual $2-3$ kilosecond observations for time monitoring and aggregate observations to obtain a high signal to noise spectrum. The aggregate {\em Swift} spectrum and folded model are shown in Figure~\ref{fig:xspectrum}, where the three components have emission measures of $7.3 \times 10^{52}$, $2.3\times 10^{52}$, and  $27\times 10^{52}$ cm$^{-3}$, respectively. 
  
For the individual {\em Swift} observations, we integrate the model X-ray spectra from 1 to 10 keV to obtain the total luminosity per observation assuming a distance of $d=161$~pc  \citep{gaia}. We do not remove the absorption component in calculating the X-ray luminosity because it is likely associated with the source itself \citep{gunther2010}. The {\em Swift} observed X-ray luminosity varied between $L_{\rm X} = (9.2\pm1.8)\times10^{29}$ erg s$^{-1}$ up to $L_{\rm X} = (4.5\pm0.9)\times10^{30}$ erg s$^{-1}$.  

\subsection{Observed Features}

\begin{figure}[ht!]
\begin{centering}
\includegraphics[width=0.46\textwidth]{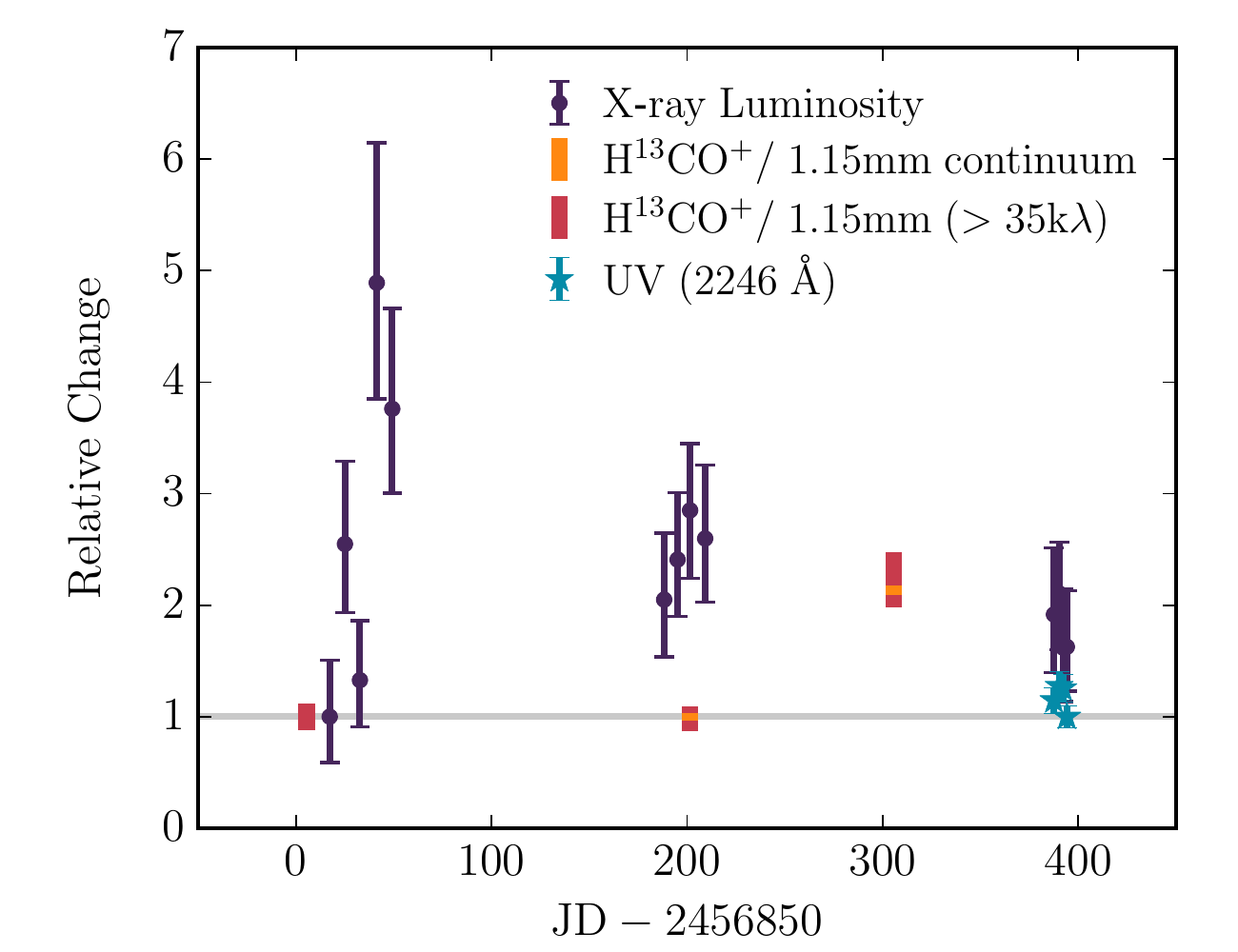}
\caption{Normalized time evolution for IM Lup's X-ray, UV, and the H$^{13}$CO$^+$ line-to-continuum ratio. X-ray luminosity provided relative to $L_{\rm X} = 9.2\times10^{29}$ erg s$^{-1}$ (purple points). Each X-ray point corresponds to the luminosity derived from a fit to the {\it Swift} spectrum for each of the individual $2-3$ kilosecond integrations. Broadband UV luminosity in the UVM2 filter normalized to 1.49 mJy shown as yellow stars. Disk-integrated and spectrally-integrated H$^{13}$CO$^+$ line-to-continuum ratio shown by red/pink vertical bars, see Section~\ref{sec:almatest}.  \label{fig:vtime}}
\end{centering}
\end{figure}

The normalized ALMA line-to-continuum ratios and {\em Swift} X-ray luminosities versus time are plotted in Figure~\ref{fig:vtime}. 
The most striking feature in the ALMA data is the H$^{13}$CO$^+$ $J=3-2$ enhancement with no change to the continuum in the spectral window containing the line shown in Figure~\ref{fig:alma}. 
The continuum traces the midplane temperature and bulk dust properties, suggesting that there was no associated change in the disk physical structure (bulk density, temperature) during the period of elevated H$^{13}$CO$^+$ $J=3-2$ flux.  Based upon additional observations of HC$^{18}$O$^+$ $J = 3-2$ to be presented in a follow-up analysis, we find the H$^{13}$CO$^+$ line is consistent with being optically thin and hence attribute changes in flux to changes in the abundance.  
The increase appears uniform on both the red- and blue-shifted sides of the disk. Between the latter two epochs the line-to-continuum ratio increased by a factor of 2.1, or $28\sigma$.

Regarding IM Lup's X-ray emission, earlier observations reported in \citet{gunther2010} found minimal change (less than a factor of two) between four observations taken between 2005 -- 2009. The {\it Swift} data (Figure~\ref{fig:vtime}) show the star has gone through more active periods, especially near MJD 2456860, consistent with typical behavior of X-ray active T Tauri stars. 
We note that the X-ray flux measurements are likely affected by multiple flares. The X-ray luminosity remained elevated during the other observations. It is important to emphasize that the stellar variability is not easily predictable and occurs over short timescales, and as a result we do not have much direct information about the 200 day stretch between the two {\it Swift} campaigns that bookend the elevated H$^{13}$CO$^+$ $J=3-2$ observation. However, the X-ray observations do point to an active, time-variable, and X-ray bright young star with increases in X-ray luminosity of at least a factor of five and likely higher if we did not catch each flare at peak luminosity.

\section{X-ray Flare Chemical Models}\label{sec:physmod}

Based upon the confirmed presence of X-ray variability and the known sensitivity of HCO$^+$ to the magnitude of X-ray irradiation \citep{cleeves2014par}, we have constructed chemical models with time-dependent X-ray ionization to investigate the impact of flares on the time evolution of HCO$^+$ as a possible explanation for the variable H$^{13}$CO$^{+}$. The key reaction governing HCO$^+$ formation in the disk atmosphere is
\begin{equation}
\rm H_3^+ + CO \rightarrow HCO^+ + H_2,
\end{equation}
where $\rm H_3^+$ is produced by X-ray ionization of $\rm H_2$ in the surface layers of the disk. The subsequent destruction is primarily dissociative recombination with electrons and charged grains, e.g.,
\begin{equation}
\rm HCO^+ + e^- \rightarrow H + CO, 
\end{equation}
where the electrons originate from UV photo-ionization of atomic carbon, so are not strongly impacted by the changing X-ray flux directly. 

To model the full time evolving physical and chemical conditions, we use a simplified version of the \citet{fogel2011} chemical code designed to model point locations within the disk. The reaction network includes 5956 reactions and 644 species including those above, initially drawn from the OSU gas-phase network of \citet{smith2004} with simple grain surface chemistry as described in \citet{cleeves2014par}. 

The input physical parameters are drawn from the IM Lup model presented in \citet{cleeves2016b}. The present paper uses this model basis to explore particular locations where HCO$^+$ is expected to be abundant \citep[normalized heights above the midplane, $z$, of $z/r$ of 0.4 and 0.5;][]{cleeves2014par}. For this initial exploration, we explore the chemical evolution at three radial locations, $r = $ 25 AU, 50 AU, and 100 AU, at both vertical ($z/r$) positions.  The input model parameters include the gas density, $\rho_{\rm gas}$, gas/dust temperature $T_{\rm gas}$ and $T_{\rm dust}$, the vertical gas column density $N_{\rm H_2}$, the local UV flux, and the X-ray ionization rate pre-flare $\zeta_{\rm X}$ (see below), see Table~\ref{tab:mod}. 

The baseline $\zeta_{\rm X}$ in Table~\ref{tab:mod} is calculated assuming the smoothed X-ray model (Figure~\ref{fig:xspectrum}) as the central X-ray source (i.e., the star), which is propagated through the IM Lup disk model with the Monte Carlo radiation transfer code of \citet{bethell2011u} using the \citet{bethell2011x} X-ray absorption cross sections. The radiation transfer is calculated at X-ray energies of 1, 2, 4, 6, 8, 11, 15, and 20 keV, to capture the changing spectral shape as a function of position in the disk. The local X-ray spectrum is extracted at each of the six points in Table~\ref{tab:mod}, with which the ionization rate $\zeta_{\rm X}$ is computed assuming the H$_2$ and helium cross sections of \citet{yan1998} and integrated over photon energy, including secondary ionizations.

We allow the incident X-ray flux to vary in time using a rapid exponential rise followed by a slower exponential decay.  The $e$-folding timescale for T Tauri star rise/decay phases span a wide range \citep[e.g.,][]{imanishi2003,favata2005,getman2008}, where rise times {($t_r$)} are typically rapid, $\sim$ hours, while the time for decay {($t_d$)} can last a few hours or even days. The primary challenge is that these calculations require $\sim 30$ minute resolution to capture the effect of individual flares, but must also cover astrophysically relevant timescales. To address this issue, the chemistry is first evolved forward for 0.5 Myr with a non-variable X-ray flux to reach ``typical'' bulk disk chemical abundances. We confirm the model has reached steady state by examining no further chemical evolution in HCO$^+$ from 0.5 to 1 Myr.  We then ``switch on'' X-ray flares and compute the chemistry at 30 minute intervals over a two-month period. 

The flaring behavior is applied in the code by directly increasing the local ionization rate of H$_2$ and helium.  For these initial simple models, we do not consider changes in the X-ray spectral shape (i.e., hardening). While the ionization cross section is similar for both soft and hard X-rays, hard X-rays have an advantage in that they are able to penetrate further into the disk. Thus a hardened spectrum can ionize denser material, perhaps requiring a less energetic flare to achieve the same increase in column density of HCO$^+$. We do not have constraints on the shape of IM Lup's X-ray spectrum during the ALMA event, and so we do not attempt to model these effects, but will explore the predicted behavior including hardening in a follow-up paper.

Given the magnitude of the change in H$^{13}$CO$^+$, we focus on the effects of strong flares, where the steady state ion abundance typically is proportional to the square root of the X-ray ionization rate. We consider three types of flares as possible explanations for the observed variability: 1) a $35\times$ change in X-ray luminosity with $t_r=3$ hours and $t_d=5$ hours and a net energy release of $E_{\rm flare} = 9.3\times10^{35}$ ergs; 2) a $80\times$ flare with $t_r=1$ hours and $t_d=3$ hours and $E_{\rm flare} = 1.1\times10^{36}$ ergs, and 3) a $10\times$ flare with $t_r=10$ hours and $t_d=30$ hours and $E_{\rm flare} = 1.3\times10^{36}$ ergs.

\begin{deluxetable*}{ccccccccc}[hb!]
\tablecolumns{9}
\tablewidth{0pt}
\tablecaption{Chemical Model Parameters\label{tab:mod}}
\tabletypesize{\footnotesize}
\tablehead{$r$ & $z/r$ & $\rho_{\rm gas}$ & $T_{\rm gas}$  & $T_{\rm dust}$   & Vertical $N_{\rm H_2}$ & UV Flux &Quiescent $\zeta_{\rm X}$ & Quiescent $\chi$(HCO$^+$)   \\
              (AU)      &      &  (g cm$^{-3}$)       &     (K)                    &  (K)                  &  (cm$^{-2}$) & (phot cm$^{-2}$ s$^{-1}$) & (s$^{-1}$ H$_2$$^{-1}$) & at 0.5 Myr (per total H) }
\startdata
25 & 0.4 & $3.9\times10^{-16}$ & 113 & 95 & $1.1\times10^{22}$& $2.3\times10^{11}$ &  $4.1\times10^{-14}$ & $3.5\times10^{-9}$ \\
25 & 0.5 & $1.6\times10^{-17}$ & 228 & 97 & $2.6\times10^{20}$& $4.9\times10^{11}$ &  $2.0\times10^{-13}$ & $2.3\times10^{-9}$\\

50 & 0.4 & $1.8\times10^{-16}$ & 87 & 77 & $5.1\times10^{21}$ & $1.4\times10^{10}$ &  $2.3\times10^{-15}$ & $1.9\times10^{-9}$\\
50 & 0.5 & $1.2\times10^{-17}$ & 125 & 78 &  $1.3\times10^{20}$ & $1.1\times10^{11}$ &$4.7\times10^{-14}$ & $2.0\times10^{-9}$ \\

100 & 0.4 & $9.1\times10^{-17}$ & 62 & 59 & $7.3\times10^{21}$ & $1.9\times10^{09}$ &  $3.1\times10^{-16}$ & $1.0\times10^{-9} $\\
100 & 0.5 & $9.7\times10^{-18}$ & 76 & 61 &  $1.8\times10^{20}$ & $2.4\times10^{10}$ &$8.8\times10^{-15}$ & $1.5\times10^{-9} $

\enddata
\end{deluxetable*}

\section{Model Results}

\begin{figure*}
\begin{centering}
\includegraphics[width=0.86\textwidth]{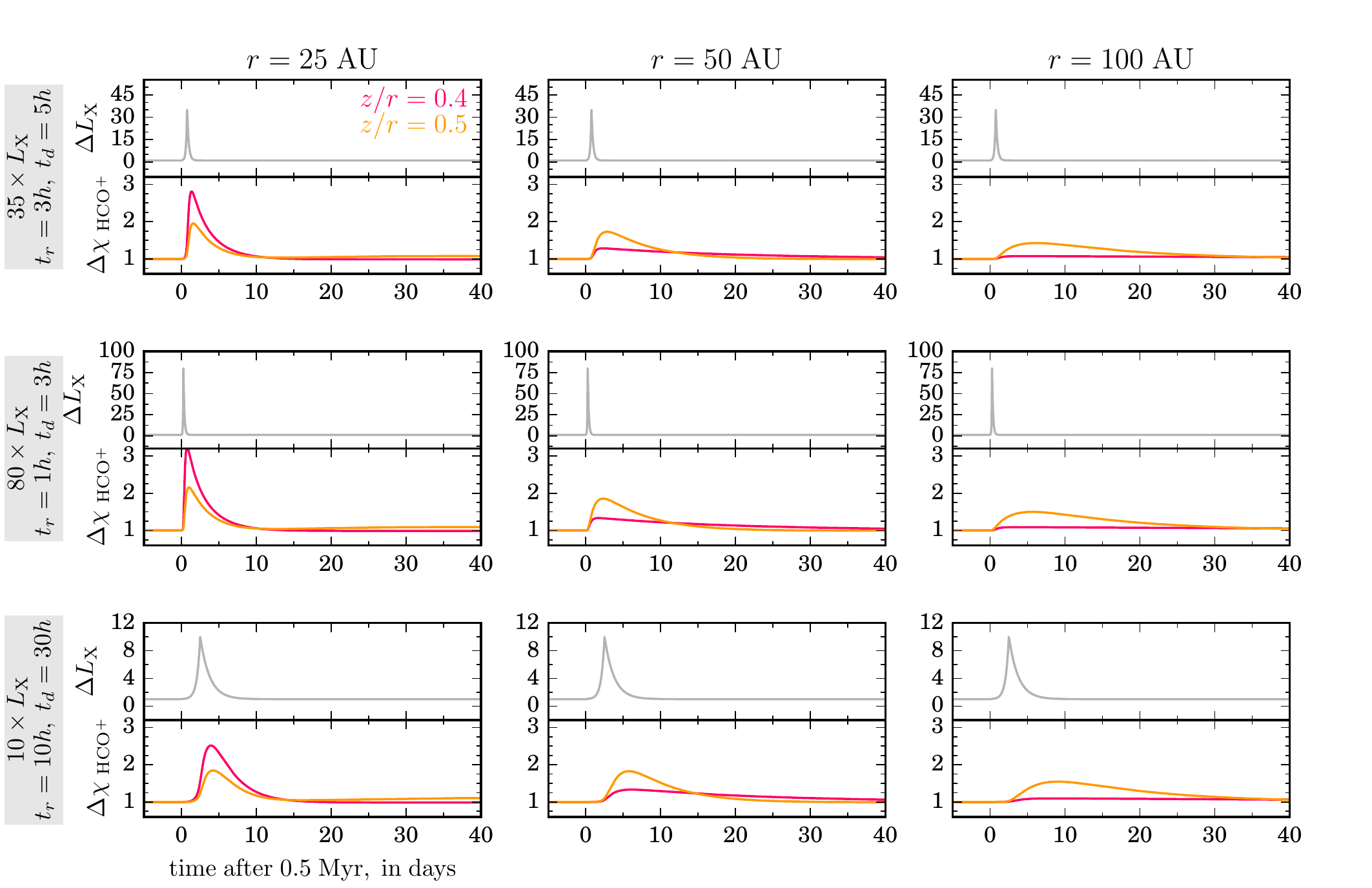} 
\caption{Time evolution of X-rays (top panels, grey) and HCO$^+$ (bottom panels, color) for three types of flares at different radial locations. Colors indicate the vertical location as labeled at the top.%
\label{fig:ionchem}}
\end{centering}
\end{figure*}

Figure~\ref{fig:ionchem} presents the HCO$^+$ abundance versus time for different X-ray models. 
We find that the H$_3^+$ abundances respond almost instantaneously to X-ray flares, followed by HCO$^+$. We expect H$^{13}$CO$^+$ and HCO$^+$ to have similar time evolution, based upon the proportionality of the doubling timescale to the HCO$^+$/CO abundance ratio: 
\begin{equation}
t_{\rm double} = \frac{\chi ({\rm HCO^+})}{\kappa n_{\rm H} \chi ({\rm CO}) \chi ({\rm H_3^+})},
\end{equation}
where $\chi$ indicates the abundance of the particular species relative to total H number density, $n_{\rm H}$, and $\kappa$ is the rate coefficient for the $\rm H_3^+ + CO \rightarrow HCO^+ + H_2$ reaction. As long as there is no strong fractionation difference between CO and HCO$^+$, both HCO$^+$ and H$^{13}$CO$^+$ evolve similarly in magnitude and time.

For the $R= 50$ and 100 AU models, $z/r = 0.5$ has the strongest response to flares (and also the highest bulk HCO$^+$ abundance), but the high electron fraction in this layer limits the effect's duration ($\sim$10 days). At $z/r = 0.4$, X-rays are two orders of magnitude more attenuated (hence a reduced HCO$^+$ response) but the lower electron density allows the effect to persist longer, lasting more than a month. For the $r=25$ AU locations, the HCO$^+$ abundance increases substantially in response to the flare due to the overall higher ionization rates present; however, the $z/r = 0.5$ position has a weaker response than $z/r = 0.4$ because it is damped by the especially high electron abundance ($\chi_e = 10^{-6}$ per H) at this position. 
 
For all flare types, the timescale for HCO$^+$ decay increases with distance from the star due to the decreasing electron volume density (and hence longer recombination times). Thus by measuring the timescale of HCO$^+$ disappearance post-flare we can estimate the local electron density. We can relate the decay timescale of HCO$^+$ to the electron density using Eq. (1) and (2):
\begin{equation}
\frac{d n_{\rm HCO^+}}{dt}  = \kappa n_{\rm CO} n_{\rm H_3^+} - \alpha n_e n_{\rm HCO^+},
\end{equation}
where $\kappa$ is the rate coefficient for the reaction in Eq. (1), $\alpha$ is the recombination rate, $\alpha = 2.8\times10^{-7} (T_{\rm gas} / 300~{\rm K})^{-0.69}$~cm$^{3}$ s$^{-1}$ \citep{amano1990}, and $n$ are the number densities of the respective species. During the post flare decay, the destruction term dominates the formation and so Eq. (4) has a simple solution:
\begin{equation}
n_{\rm HCO^+} = n_{\rm HCO^+ (peak)} ~ e^{-t / \alpha n_e},
\end{equation}
such that the time for post-flare HCO$^+$ abundance to halve is 
\begin{equation}
t_{\rm half} = -\frac{\ln{(1/2)}}{ \alpha n_e}.
\end{equation} 
Thus by measuring $t_{\rm half}$, $n_e$ can be estimated. For example, the model at $r=50$~AU, $z/r=0.5$ reaches half peak HCO$^+$ abundance at $\sim 5$~days. With Eq. (6) and assuming $T_{\rm gas} = 100$ K, $n_e$ is estimated to be $\sim2.7$~cm$^{-3}$ from these equations, where the value from the detailed chemical model is 3.3~cm$^{-3}$. In practice, lack of temperature information introduces a factor of a few uncertainty in the derived $n_e$.

\section{Discussion}\label{sec:discussion}

We report the detection of significant variability in the  H$^{13}$CO$^+$ $J=3-2$ line emission in the IM Lup disk relative to the dust continuum, which remains unchanged within the uncertainty. Our models demonstrate one possible explanation for the observed behavior is enhanced ion chemistry during an X-ray flare.
Impulsive strong flares ($35\times$ or $80\times$ $L_{\rm X}$) or weaker ($10\times$ $L_{\rm X}$) but long-duration flares are sufficient to explain the increase in H$^{13}$CO$^+$ $J=3-2$, where the net change is related to the enhancement of H$_3^+$ from X-rays and the length of time the reaction can proceed (see Eq.~4). A closely clustered series of short $10\times L_{\rm X}$ flares can also mimic a long duration event and reproduce the observed enhancement. 

While the present observations provide the first robust evidence of submillimeter variability, earlier work of \citet{thi2004} noted that observations reported in \citet{zadelhoff2001} measured HCO$^+$ $J=4-3$ fluxes toward TW~Hya in 1999 that were a factor of three fainter compared to earlier line fluxes reported in \citet{kastner1997} from 1995 -- 1996, and \citet{thi2004}'s own later measurements in 2004. During this same time frame, measurements of the neutral species CO, HCN, and CN at different epochs had fluxes in agreement. The authors suggest two possibilities: 1) there may be a systematic issue with the unresolved data or 2) the variability is robust. Our data indicates that the latter is indeed a viable interpretation.

Indirect evidence of flare-driven chemistry was also found via models of ionization chemistry in TW Hya, where HCO$^+$ was better fit with a hardened (i.e., flaring) X-ray spectrum compared to TW Hya's ``quiescent'' X-ray spectrum \citep{cleeves2015tw}. Essentially the HCO$^+$ abundance was in an elevated state where the timescales between flares were shorter than the time to dissociatively recombine and remove HCO$^+$. We examined the effect of frequent, factor of two increases in X-ray luminosity every two days and find that indeed the HCO$^+$ abundance can stay in a constant $\sim10\%$ elevated state compared to a non-flaring model.

These types of observations open a new window into our understanding of disk ionization chemistry, which is more dynamic than previously assumed. In the future, time-resolved simultaneous X-ray and submillimeter observations during an X-ray flare may prove a useful new tool to understand disk physics. With spatially resolved H$^{13}$CO$^+$ imaging, we may be able to watch the propagation of a stellar flare across the face of a disk where the light crossing time over 100 AU scales is $\sim 14$ hours. For flares occurring near the poles, the emission would shine outward symmetrically in azimuth over the entire disk. For flares originating at lower stellar latitudes (away from the poles), the flare may illuminate only a fraction of the disk azimuthal area.

Following the initial propagation of the flare, the H$^{13}$CO$^+$ signature should fade over week to month-long timescales, depending on the local electron density. Thus this technique provides a new method to estimate the disk's spatial ionization fraction. This parameter is key in our understanding of disk accretion physics, thermal structure, and disk chemistry. One would need only slightly deeper ALMA observations ($\sim1$ hour) to produced resolved images, repeated over the duration of the light travel time ($\sim1$ day total), to follow the immediate propagation of X-ray ``echoes'' across the surface of the disk. During the decay phase, the H$^{13}$CO$^+$ could be imaged every $\sim1-3$~days to directly measure the timescale for recombination with electrons. Such observations are readily within our current technical capabilities, and could be complimented with simultaneous observations of additional molecules to explore the extent of variable ionization chemistry in disks.

\acknowledgements{{\it Acknowledgements:} The authors are grateful to the anonymous referees whose comments improved the manuscript. The authors are also appreciative of the NRAO ALMA Helpdesk in assisting with the data quality assessment. This paper makes use of the following ALMA data: ADS/JAO.ALMA\#2013.00694 and ADS/JAO.ALMA\#2013.1.00226.S. ALMA is a partnership of ESO (representing its member states), NSF (USA) and NINS (Japan), together with NRC (Canada) and NSC and ASIAA (Taiwan), in cooperation with the Republic of Chile. The Joint ALMA Observatory is operated by ESO, AUI/NRAO and NAOJ. The National Radio Astronomy Observatory is a facility of the National Science Foundation operated under cooperative agreement by Associated Universities, Inc. This work made use of data supplied by the UK Swift Science Data Centre at the University of Leicester and we acknowledge the use of public data from the Swift data archive. LIC acknowledges the support of NASA through Hubble Fellowship grant HST-HF2-51356.001-A awarded by the Space Telescope Science Institute, which is operated by the Association of Universities for Research in Astronomy, Inc., for NASA, under contract NAS 5-26555. EAB acknowledges support from National Science Foundation grant AST-1514670 and AST-1344133 (INSPIRE) along with NASA XRP grant NNX16AB48G. KI\"O also acknowledges funding through a Packard Fellowship for Science and Engineering from the David and Lucile Packard Foundation. RAL gratefully acknowledges support from the NRAO Student Observing Support Program.  }


\end{document}